\documentclass[useAMS,usenatbib]{mn2e}

\usepackage{graphics,graphicx}
\usepackage{epstopdf}
\usepackage{epsfig}

\usepackage[usenames]{color}

\title[The mass and orbit of LS~5039]{The gamma-ray binary LS~5039: mass and orbit constraints from {\em MOST} observations\thanks{Based on data from the {\em MOST} satellite, a
Canadian Space Agency mission, jointly operated by
Microsat Systems Canada Inc. (MSCI, formerly the space division
of Dynacon Inc.), the University
of Toronto Institute for Aerospace Studies and the University of British
Columbia, with the assistance of the University of Vienna.}}

\author[Sarty et al.]{Gordon E. Sarty$^{1,2}$, Tam\'{a}s Szalai$^{3}$, L\'{a}szl\'{o} L. Kiss$^{4,5}$, Jaymie M. Matthews$^{6}$, \newauthor
Kinwah Wu$^{5,7}$, Rainer Kuschnig$^{8}$, 
David B. Guenther$^{9}$, Anthony F.J. Moffat$^{10}$, \newauthor
Slavek M. Rucinski$^{11}$, Dimitar Sasselov$^{12}$, 
Werner W. Weiss$^{8}$, Richard Huziak$^{1}$,  \newauthor
Helen M. Johnston$^{5}$, Andre Phillips$^{13}$, Michael C.B. Ashley$^{13}$\\
$^{1}$Royal Astronomical Society of Canada, Saskatoon Centre, P.O. Box 317, RPO University,
Saskatoon, SK S7N 4J8, Canada\\
$^{2}$Department of Physics and Engineering Physics, University of Saskatchewan, 
Saskatoon, SK S7N 5E2, Canada\\
$^{3}$Department of Optics and Quantum Electronics, University of
Szeged, D\'om t\'er 9., Szeged H-6720, Hungary\\
$^{4}$Konkoly Observatory of the Hungarian Academy of Sciences, H-1525 Budapest, P.O. Box 67, Hungary\\
$^{5}$Sydney Institute for Astronomy, School of Physics A28, 
University of Sydney, NSW 2006, Australia\\
$^{6}$Department of Physics and Astronomy, University of British Columbia, 6224
Agricultural Road, Vancouver, BC V6T 1Z1, Canada\\
$^{7}$Mullard Space Science Laboratory, University College London, Holmbury St.
Mary, Dorking, Surrey RH5 6NT\\
$^{8}$Institut f\"ur Astronomie, Universit\"at Wien, T\"urkenschanzstrasse 17, A-1180 Vienna, Austria\\
$^{9}$Department of Astronomy and Physics, Saint Mary's University, Halifax, N.S., B3H 3C3, Canada\\
$^{10}$Observatoire Astronomique du Mont M\'{e}gantic, D\'epartment de Physique,
Universit\'e de Montr\'eal C. P. 6128,\\ \ \ Succursale: Centre-Ville, Montr\`{e}al,
QC H3C 3J7, Canada\\
$^{11}$Department of Astronomy and Astrophysics, University of Toronto, Toronto, ON M5S 3H4, Canada\\
$^{12}$Harvard-Smithsonian Center for Astrophysics, 60 Garden Street, Cambridge, MA 102138, USA\\
$^{13}$School of Physics, Department of Astrophysics and Optics, University of New South Wales, Sydney, NSW 2052, 
Australia}

\begin{document}

\date{in original form 2010}

\pagerange{\pageref{firstpage}--\pageref{lastpage}} \pubyear{2010}

\maketitle

\label{firstpage}

\begin{abstract}

The results of a coordinated space-based photometric and ground-based spectroscopic 
  observing campaign on the enigmatic $\gamma$-ray binary LS~5039 are reported. 
Sixteen days of observations from the {\em MOST} satellite have been combined with high-resolution
optical  echelle spectroscopy from the 2.3m ANU Telescope in Siding Spring, Australia.  
These observations were used to measure the orbital parameters of the binary and to
  study the properties of stellar wind from the O primary. 
We found that any broad-band optical photometric variability at the orbital period  
  is below the 2 mmag level,   
  supporting the scenario that the orbital {\bf eccentricity of the system is near the 0.24$\pm$0.08 value implied by our spectroscopy, which is lower than values previously obtained by other workers. The low amplitude optical variability also implies the component masses are at the higher end of estimates based on the primary's O6.5V((f)) spectral type with a primary mass of $\sim$26 M$_{\odot}$ and a mass for the compact star of at least 1.8 M$_{\odot}$.  } 
The mass loss rate from the O primary was determined to be 3.7 to 4.8 $\times 10^{-7}$ M$_{\odot}$ yr$^{-1}$.

\end{abstract}


\section{Introduction}

LS~5039 (V479~Sct) is the optical counterpart of the peculiar X-ray source RX~J1826.2$-$1450 \citep{motch1997}. 
The system has been observed in the radio, optical/IR, UV, X-ray and $\gamma$-ray wavelengths. 
A radio counterpart has been identified in VLA observations by \citet{Marti1998}.   
The emission appeared to be persistent and was non-thermal.  
A pair of symmetric radio features were later found associated with the source (core) in VLBA observations by  
   \citet{Paredes2000, Paredes2002}, 
   which were interpreted as emission from two opposite relativistic jets.  
The system was also identified with a very high energy (VHE) $\gamma$-ray source 
  found in the {\it CGRO}/EGRET (Paredes et al. 2000) and HESS \citep{Aharonian2005a} surveys.  
The detection of MeV-GeV/TeV emission placed LS~5039/RX~J1826.2$-$1450 
   into a class of unusual high-energy objects, the $\gamma$-ray binaries. 
So far, only a handful of $\gamma$-ray binaries are known. 
The others are PSR B1259$-$63 \citep{Aharonian2005b},  
      LS I +61 303 \citep{Albert2006,Acciari2008},   
      Cygnus X-1 \citep{Albert2007},  
      Cygnus X-3 \citep{Tavani2009},  
      and the recent candidate HESS~J0632+057 \citep{Hinton2009}.  

LS~5039/RX~J1826.2$-$1450 (hereafter, referred to as LS~5039)  
   has been classified as a high-mass X-ray binary in the catalogue compiled by \cite{Liu2006}. 
The distance to the source is $\approx 2.5$kpc \citep{Casares2005}.
Its primary is a bright ($V = 11.2$) massive O star \citep{Clark2001, McSwain2001},  
   and its secondary is a compact star. 
It is still a matter of debate whether the compact star in LS~5039 is a black hole or a neutron star.  
UV and optical spectroscopy established that the primary is a O6.5V((f)) star \citep{McSwain2004}.  
The presence of  P~Cygni profiles in the UV N~{\small V}~$\lambda${1240} and C~{\small IV}~$\lambda${1550} lines 
  \citep{McSwain2004} indicates a strong wind outflow, 
  whose rate has been estimated to be $\sim {\bf 10^{-7}}$~M$_\odot$yr$^{-1}$ or even higher \citep{McSwain2004}.  
From the radial velocities of the H$\alpha$ and He lines, 
  \citet{McSwain2004} (hereafter M04) obtained an orbital period $P=(4.4267\pm0.0005)$d  
  and {\bf derived the mass function} $f(m) = (0.0017\pm0.0005)$M$_\odot$, 
  the orbital eccentricity $e=0.48\pm0.06$ 
  and $a_1 \sin i = (1.36\pm 0.13)$R$_\odot$, 
  where $a_1$ is the semi-major axis of the primary's orbit and $i$ is orbital inclination.  
They argued that the mass of the primary O star is in the range $(20-35)$M$_\odot$, 
  and the compact secondary is a neutron star which has a mass $\approx 1.4$M$_\odot$.  
\citet{Casares2005} (hereafter C05) conducted a comprehensive analysis of 
  the optical H Balmer and He{\tt I} and He{\tt II} lines  
  and obtained $P=(3.90603\pm0.00017)$d, 
  $e=0.35\pm0.04$, $f(m) = (0.0053\pm0.0009)$M$_\odot$ 
  and $a_1 \sin i = (1.42\pm 0.07)$R$_\odot$.  
They derived that the orbital inclination $i =(24.9\pm2.8)^\circ$, 
    the primary O star has a mass $M_1 = 22.9^{+3.4}_{-2.9}$M$_\odot$ 
    and the compact star has a mass $M_2 =3.7^{+1.3}_{-1.0}$M$_\odot$. 
The orbital period obtained by C05 is shorter than that obtained by M04.  
It is however consistent with the modulations observed in the X-rays 
  \citep{Bosch-Ramon2005,Takahashi2009} and $\gamma$-rays 
  at GeV \citep{Abdo2009} and TeV \citep{Aharonian2006} energies. 
\citet{Aragona2009} (hereafter A09) revisited radial velocity (RV) measurements from optical spectra, 
   confirming the 3.9d orbital period.  
Their refined orbital parameters are 
   $P=(3.90608\pm0.00010)$d, 
  $e=0.337\pm0.036$, $f(m) = (0.00261\pm0.00036)$M$_\odot$ 
  and $a_1 \sin i = (1.435\pm 0.066)$R$_\odot$.

One major question about LS~5039 is the nature of the compact object. 
{\bf The orbital parameters derived by C05 
  clearly indicate that the compact star has a mass exceeding 3.0~M$_\odot$,  
  the usually accepted upper limit of masses of neutron stars \citep{Lattimer2007},  
  implying that LS~5039 is a candidate black-hole high-mass X-ray binary. } 
Black-hole binaries with a massive O-type donor star are very rare, 
  partly because of the extremely short life-spans of such systems.     
To date, Cyg X-1 is the only known black-hole X-ray binary in the Milky Way with a massive O donor star. 
The presence of a black hole {\bf or not} in LS~5039 {\bf would therefore have} significant implications 
  not only on the formation of black-hole high-mass X-ray binaries and the population of such systems in the Milky Way, 
  but also on how very high-energy (TeV) emission  
  is produced in $\gamma$-ray binaries (see e.g. \citet{Araudo2009}).     
    
The issue of the nature of the compact star associated with LS~5039 is far from being settled.      
{\bf While the spectroscopic observations of C05 favor the black-hole scenario, 
  some workers have argued for a neutron-star scenario 
  involving a non-accreting young pulsar \citep{Martocchia2005,Dubus2006,Sierpowska2007,Cerutti2008}. }
Observationally, 
  \citet{Ribo2008} found that changes in the {\em mas} morphology of the radio images 
  were difficult to reconcile with the micro-quasar (black-hole binary) scenario 
  in which the radio emission originates from expanding plasmons.  
The situation is further complicated by the fact that  
  the temporal and spectral behaviour of the X-ray emission  
  \citep{Bosch-Ramon2007,Bosch-Ramon2009,Takahashi2009,Kishishita2009} 
  and properties of the TeV emission \citep{Bosch-Ramon2008,Khangulyan2008,Abdo2009} 
  indicate that the high-energy radiation  
  might originate from regions far outside the binary orbit of LS~5039.  

We have been observing LS~5039 photometrically with ground-based telescopes at optical and near IR wavelengths 
to search for orbital modulation, but  
orbital variations have not been detected in those data (to be reported elsewhere). 
However, we have detected variations at the 20 mmag level, especially in the $I_{C}$-band.  
This prompted us to take a closer look at the optical photometric variations of LS~5039 
  from space with the Canadian Microvariability and Oscillations of Stars ({\em MOST}\ ) satellite. 
Our {\em MOST} observations were made in July 2009, 
  simultaneously with ground-based optical spectroscopic observations 
  from the Australian National University (ANU) 2.3m Telescope at Siding Spring Observatory, Australia.   
Here we report on the main findings, 
  especially those concerning the masses of the component stars in LS~5039 
  and the nature of the compact star.  

This paper is organized as follows. 
In Section \ref{two} 
   we describe the photometric observations with the  {\em MOST} space telescope 
   and the spectroscopic observations with the ANU 2.3m Telescope, which were made simultaneously. 
In Section \ref{three} we describe the data analysis, 
  present the results on the system's orbital and other parameters,  
  and discuss the implications of the nature of the compact object  
  and the properties of the wind/outflow from the primary star. 
The conclusions are summarized in Section \ref{four}. 

\section{Simultaneous photometric and spectroscopic observations}\label{two}

\begin{figure}
\begin{center}
\leavevmode

\includegraphics[width=8.2cm]{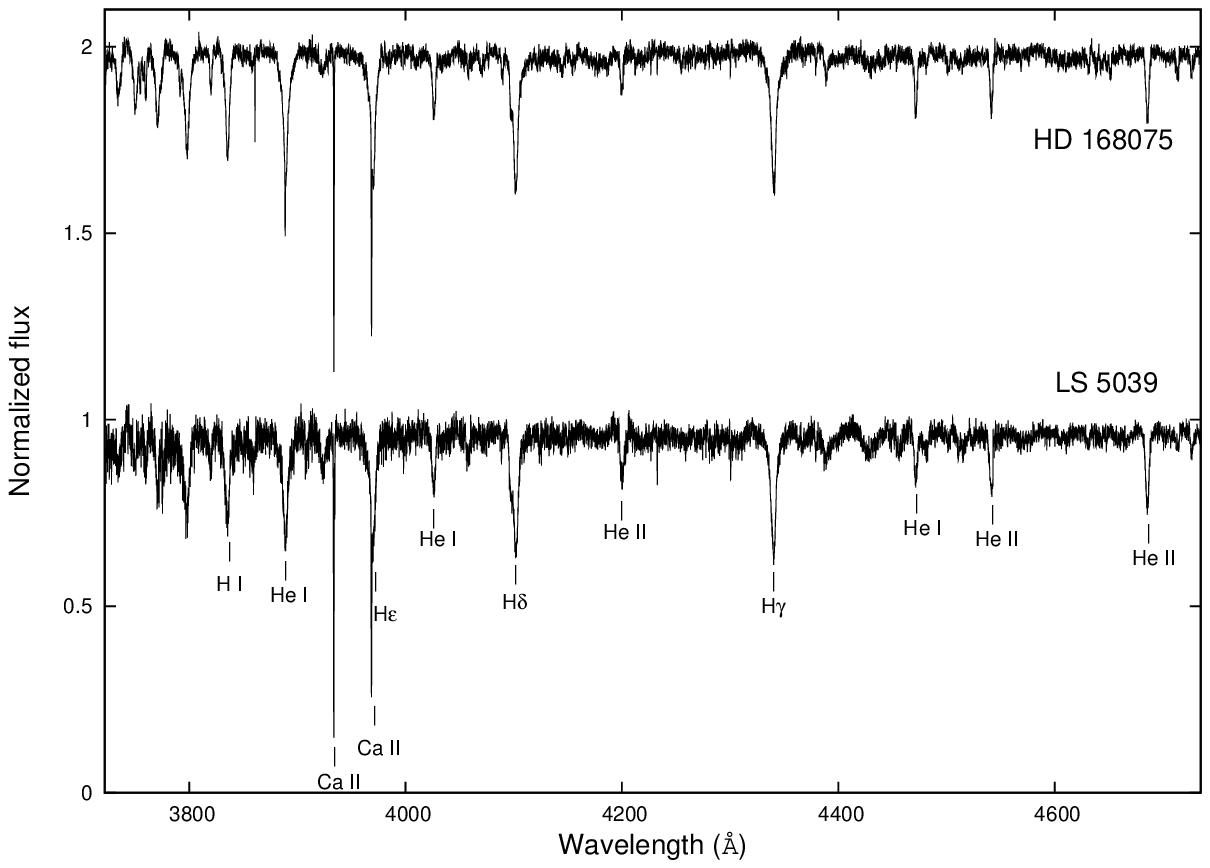}
\includegraphics[width=8.2cm]{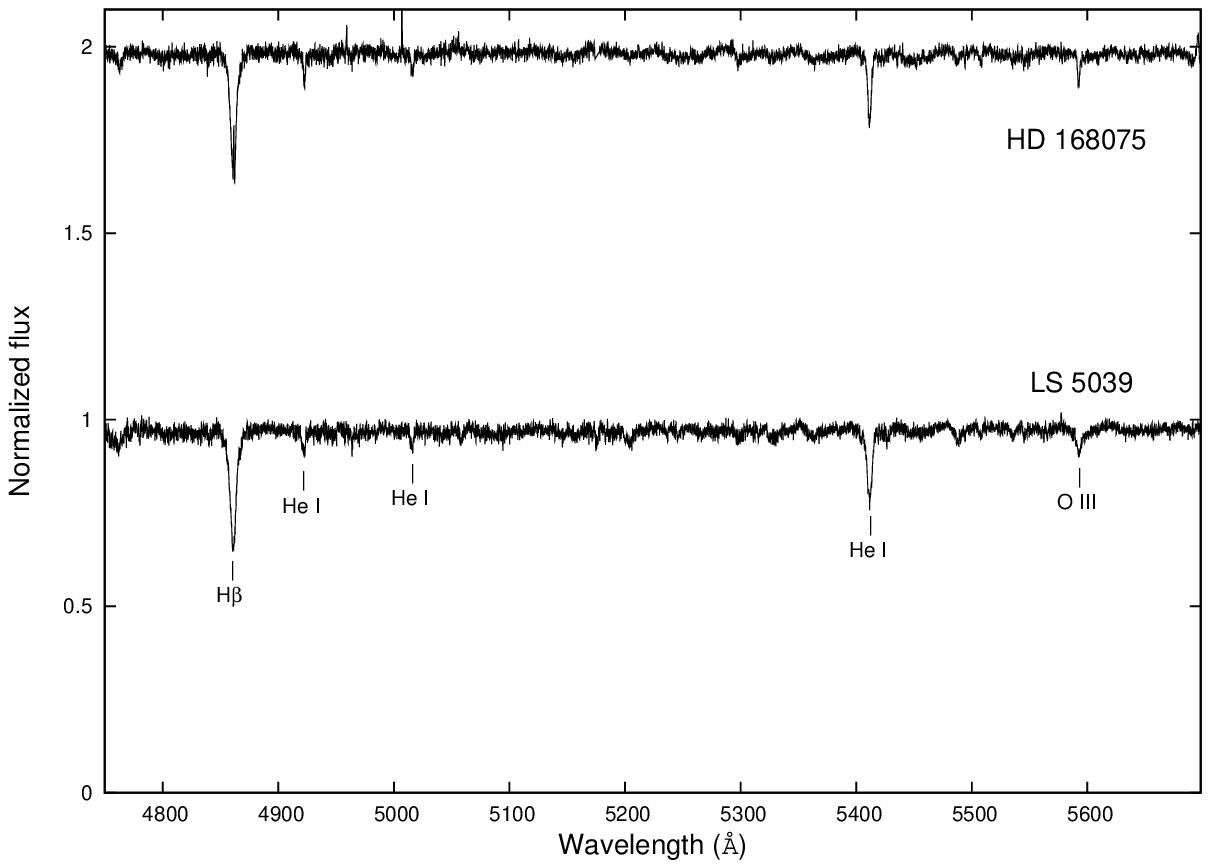}
\includegraphics[width=8.2cm]{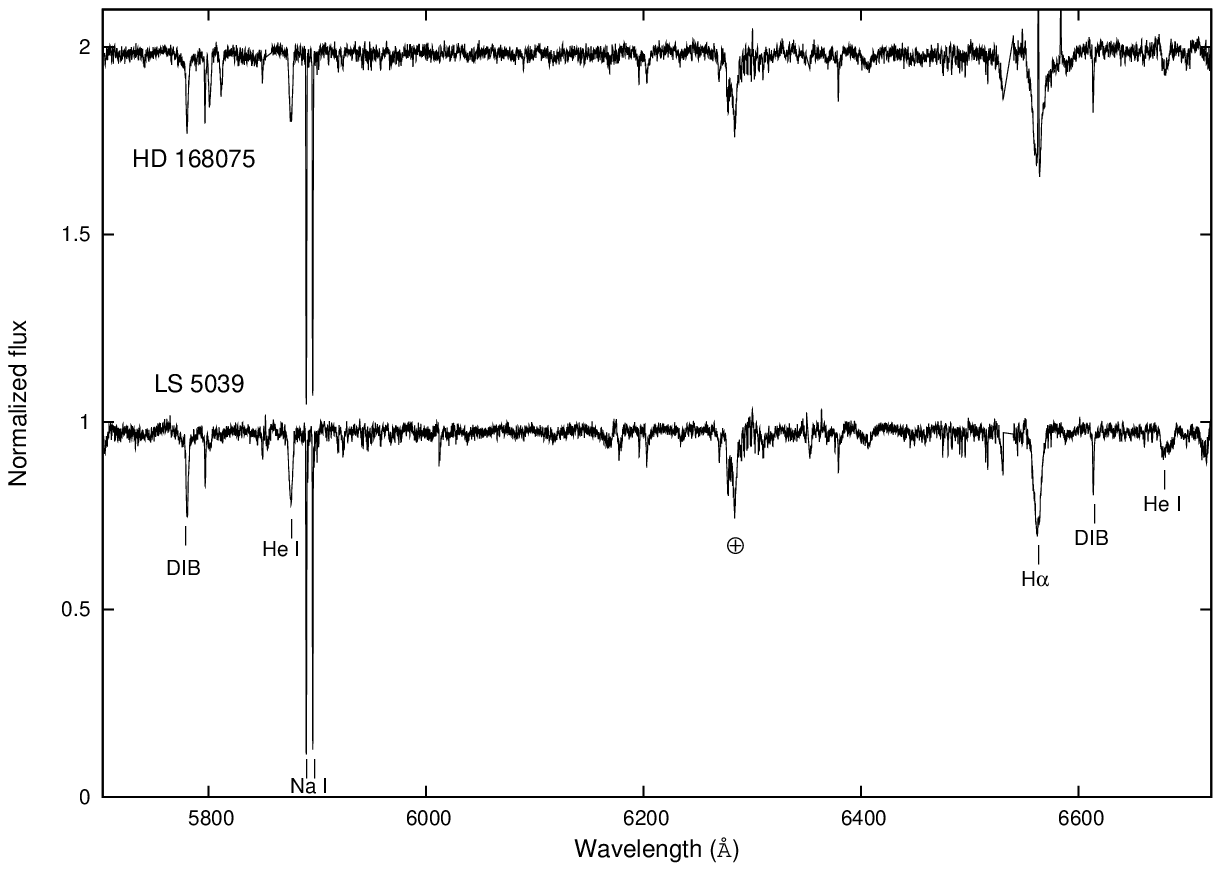}
\end{center}
\caption{Continuum normalized spectra of LS~5039, classified as type O6.5V((f)), and the O6-7V((f)) spectral template star HD 168075.}
\label{longsp}
\end{figure}

\subsection{{\em MOST} photometry}

The {\em MOST} microsatellite \citep{Matthews1999,Walker2003} houses a 15-cm Rumak-Maksutov telescope feeding a
CCD photometer through a custom optical broadband (350 -- 750 nm)
filter.  From its polar Sun-synchronous orbit (altitude = 820 km;
orbital period = 101 min), it has a Continuous Viewing Zone (CVZ)
about $54^{\circ}$ wide within which it can monitor target fields
for up to 2 months without interruption.  Targets brighter than
V $\sim$ 6 are observed in Fabry Imaging mode; fainter targets (like
LS 5039) are observed in Direct Imaging mode, similar to standard
CCD photometry with a ground-based instrument.  The photometry is
non-differential, but given the orbit, thermal and design
characteristics of {\em MOST}, experience has shown that it is a very
photometrically stable platform even over long timescales (with
repeatability of the mean instrumental flux from a non-variable
target of the brightness of LS 5039 to within about 1 mmag).

The science exposures on the {\em MOST} Science CCD take place at the
same time as the guide star exposures for satellite pointing.  In
the case of the LS 5039 observations, the guide star exposure
time (and hence the science target exposure time) was 3.03 s.
To build up Signal-to-Noise Ratio (SNR), the science exposures were co-added on board the
satellite in `stacks' of 10 exposures.  Stacks (each 30.3 s of
total integration) were downloaded from the satellite consecutively
(with no dead time between stacks), giving a sampling rate of about
twice per minute.

The LS 5039 field lies outside the {\em MOST} CVZ so it could not be
monitored continuously.  This field alternated with another {\em MOST}
Primary Science Target field (the Wolf-Rayet star WR 113) during
each {\em MOST} satellite orbit.  Typically, we monitored LS 5039 for
about $70\%$ of every second {\em MOST} orbit (or about 70 of every second interval of 101 min) between 2009 July 7 -- 23.

Due to the observing season and the location of the LS 5039 field
relative to the illuminated limb of the Earth, scattered earthshine
was high during the beginning and end of each `visit' to the LS 5039
field.  We truncated about 15 -- 20 min of each orbit from the
original data to preserve the highest photometric precision.  We
also filtered outliers caused by cosmic ray hits, especially during
passages of the satellite through the South Atlantic Anomaly (SAA).
However, the resulting light curve has an effective duty cycle of
about 50\% (with gaps at intervals of about 101 min) and it still
sampled the important timescales in the LS 5039 system thoroughly.

\subsection{Echelle spectroscopy}

The spectroscopic observations were carried out on four nights between 2009 July 8 -- 11 (during the {\em MOST} observations) 
and
on three nights between August 
1 -- 3, using
   the ANU 2.3m Telescope with an echelle spectrograph. In 
total, 118 spectra were obtained that cover almost 40 hours with nearly uniform sampling of the whole orbit. 
The integration times were between 900 -- 1200 s and the spectra covered the whole visual
range between 3900 \AA\ and 6720 \AA. The nominal spectral resolving power was $\lambda/\Delta \lambda\approx$ 23,000 at 
the 
H$\alpha$ line, with a typical SNR per extracted pixel of about 100 for combined spectra that represented one-hour integrations.

All data were reduced with standard {\small IRAF}\footnote{{\small IRAF} is distributed by the National Optical Astronomy 
Observatories, which are operated by the Association of Universities for Research in Astronomy, Inc., under co-operative 
agreement with the National Science Foundation.} tasks, including bias and flat-field corrections, cosmic ray removal, 
extraction of the 27 individual orders of the echelle spectra, wavelength calibration, and continuum normalization. ThAr 
spectral lamp exposures were regularly taken before and after every object spectrum to monitor the wavelength shifts of 
the CCD spectra. We also obtained spectra for the telluric standard HD 177724 and the O6-7V((f)) spectral template
HD 168075 \citep{Dufton2006}. Typical continuum-normalized spectra for LS~5039 and HD 168075 are shown in Fig.~\ref{longsp}.

\section{Data analysis and results}\label{three}

\begin{figure}
\begin{center}
\leavevmode

\includegraphics[width=8.5cm]{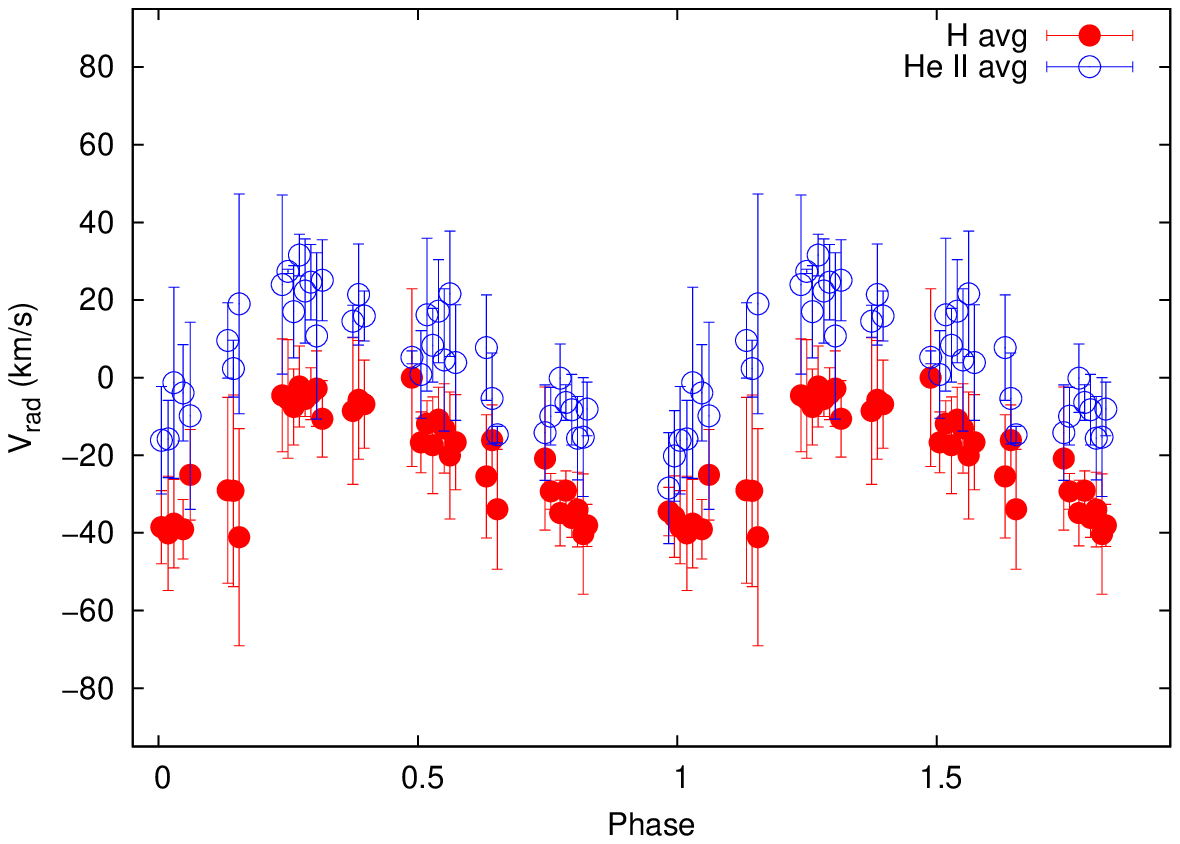}
\includegraphics[width=8.5cm]{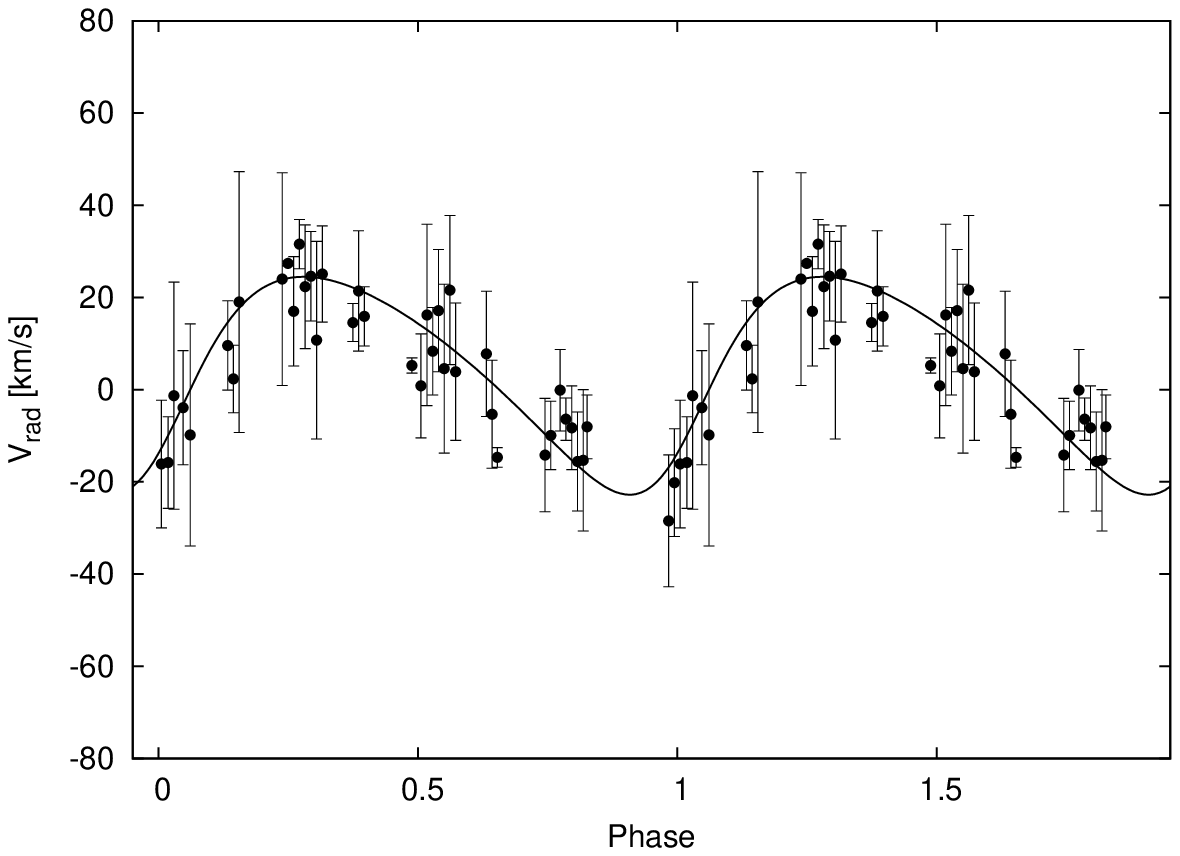}

\end{center}
\caption{{\it{Top}:} Radial velocities based on H Balmer and He II lines. {\it{Bottom}}: The best-fitting curve to radial 
velocities of He II lines.}
\label{rvfit}
\end{figure}

\begin{table*}
\caption{Orbital parameters of LS 5039.}
\begin{center}
\begin{tabular}{|l|l|l||l|}
\hline
Parameter & C05 (He \begin{small}II\end{small})& A09 & This paper (He \begin{small}II\end{small})\\
\hline
\hline
$T_0$ (HJD$-$2450000) & 1943.09 $\pm$ 0.10 & 2825.99 $\pm$ 0.05 & 5017.08 $\pm$ 0.06\\ 
$P_{\rm orb}$ (d) & 3.90603 & 3.90608 & 3.906 (adopted)\\
$e$ & 0.35 $\pm$ 0.04 & 0.34 $\pm$ 0.04 & 0.24 $\pm$ 0.08\\
$\omega$ ($^{\circ}$) & 225.8 $\pm$ 3.3 & 236.0 $\pm$ 5.8 & 237.3 $\pm$ 21.8\\
$V_{\gamma}$ (km s$^{-1}$) & 17.2 $\pm$ 0.7 & 4.0 $\pm$ 0.3 & 3.9 $\pm$ 1.3\\
$K_1$ (km s$^{-1}$) & 25.2 $\pm$ 1.4 & 19.7 $\pm$ 0.9 & 23.6 $\pm$ 4.0\\
$a_1$ sin $i$ (R$_{\odot}$) & 1.82 $\pm$ 0.10 & 1.44 $\pm$ 0.07 & 1.77 $\pm$ 0.15\\
$f(m)$ (M$_{\odot}$) & 0.0053 $\pm$ 0.0009 & 0.0026 $\pm$ 0.0004 & 0.0049 $\pm$ 0.0006\\
RMS of fit (km s$^{-1}$) & 9.1 & 7.1 & 6.2\\
\hline
\end{tabular}
\end{center}
\label{orbitpar}
\end{table*}

\subsection{Radial velocities}\label{RV}

To measure radial velocities we first generated one hour long averaged spectra to get higher SNR -- one hour corresponds
to 0.01 orbital phase, hence negligible phase smearing appears in the phased radial velocity data. Radial 
velocities of the H \begin{small}I\end{small}, He \begin{small}I\end{small} and He \begin{small}II\end{small} lines were 
determined by fitting two-component functions (a concentric sum of Gaussian and Lorentzian functions) and calculating the shift of the
centroids to laboratory wavelengths. To check our method we also determined the velocities of some interstellar lines 
(Ca \begin{small}II\end{small} K, Na \begin{small}I\end{small} D, DIB $\lambda$5780, DIB $\lambda$6613) for each night. We 
applied heliocentric corrections to the radial velocity for each line at each time.  
The data were phased with an orbital period of 3.906 d (C05).   
The success in generating a RV curve with this phase folding  
   supports the orbital period of LS~5039 as being that obtained by C05 instead of the longer 4.4 d period obtained by M04.

To generate the final RV diagrams we used averaged velocities of the H \begin{small}I\end{small} (H$\alpha$, H$\beta$, 
H$\gamma$, H$\delta$, $\lambda$3835), He \begin{small}I\end{small} ($\lambda$4471, $\lambda$5875) and He 
\begin{small}II\end{small} ($\lambda$4200, $\lambda$4686, $\lambda$5411) lines; there are several other H and He lines 
in the
wavelength region of our spectra (e.g. the frequently used He \begin{small}II\end{small} $\lambda$4542 line), but they were 
too noisy or blended to use for velocity determination. The resulting RV diagrams are shown in Fig.~\ref{rvfit}, top.
The error bars shown in Fig.~\ref{rvfit} (typically $\pm$10 to $\pm$15 km s$^{-1}$ in magnitude) represent plus and minus one standard deviation of the measured velocities of
two to five lines at different wavelengths. The relatively large errors were caused partly by observational noise and partly 
 by the high rotational velocity of the O star ($v_{\rm rot} \sin i$ = 113 $\pm$ 8 km s$^{-1}$, C05).

We detected systematic increasing blueshifts from the He \begin{small}II\end{small} lines to the He \begin{small}I\end{small} lines to the H Balmer lines. C05 have also described this effect, but found
its degree to be smaller (they obtained an $\sim$8 km s$^{-1}$ shift between the average RVs of the He \begin{small}II\end{small} and 
H Balmer lines, as opposed to our 20 km s$^{-1}$ value). We believe that the RV shift is due to the
contamination of wind for the He \begin{small}I\end{small} and H \begin{small}I\end{small} lines (see \citet{Puls1996}), and so we did not use those two atomic line sets to constrain the orbit.

Additionally, redshifted satellite absorptions were found in the Ca \begin{small}II\end{small} K and Na 
\begin{small}I\end{small} D1 and D2 {\bf interstellar absorption} lines with a radial velocity around +60 km s$^{-1}$ (more precisely, +58.4 $\pm$ 2.2 km s$^{-1}$ by Ca K, +62.9 $\pm$ 2.3 km s$^{-1}$ by Na D1 and +61.9 $\pm$ 1.7 km s$^{-1}$ by Na D2). These satellite lines may 
belong to a formerly unknown Galactic Intermediate Velocity Cloud [IVC; see the review by \citet{Wakker1997}].

\subsection{Orbital and system parameters}\label{orbit}

In the following analysis we adopted $T_{\rm eff}$ = 39,000 $\pm$ 1000 K and log $g$ = 3.85 $\pm$ 0.10 
   for the O primary (C05). 
We measured the equivalent widths (EW) of several interstellar lines to estimate the interstellar reddening 
  (see Section \ref{starwind}  for  EW measurements of other lines). 
  Using Na \begin{small}I
\end{small} 
D1 [EW = 0.70 $\pm$ 0.02 \AA\ with the relation to reddening as per \citet{Munari1997}], DIB $\lambda$5780 and DIB $\lambda$6613 
[EW = 
0.55 $\pm$ 0.05 \AA\ and 0.18 $\pm$ 0.02 \AA, respectively with the relation to reddening as per \citet{Cox2005}] lines, we found 
$E(B-V)$ = 1.2 $\pm$ 0.1, which is in agreement with previous results [1.25 to 1.35, \citet{Ribo2002} and 1.28 $\pm$ 0.02, M04].  Based on this agreement, we adopted the values     
   $d = 2.5 \pm 0.1$ kpc, $M_{\rm O} = 22.9^{+3.4}_{-2.9}$ M$_{\odot}$ and  
   $R_{\rm O} = 9.3^{+0.7}_{-0.6}$ R$_{\odot}$ obtained by C05.  

Radial velocity curves were modelled using the 2003 version\footnote{ftp://ftp.astro.ufl.edu/pub/wilson} of the  
Wilson-Devinney
(WD) code \citep{Wilson1971,Wilson1994,Wilson2003}. We used only our own velocity points and did not use previous RV 
measurements made by others. Our data set therefore represents the highest resolution, homogenous spectral data set yet 
used to obtain an orbital solution for LS~5039.
Given the lack of X-ray eclipses we could not use the special mode in WD code developed for modeling the orbits of 
X-ray binaries. Therefore, we analysed LS~5039 as a single line spectroscopic binary without any (X-ray) light curves. This limitation
allowed us to determine only the mass function, $f(m)$, as a function of different inclination angles, $i$. We could not determine the exact values of the inclination, $i$, and the
mass ratio, $q$. However, knowledge of the primary's mass, $M_{\rm O}$, allowed us to narrow the possible parameter space. 

As described previously, we adopted the 3.906 d period as a fixed parameter during the modelling. The computed value of 
$T_{0}$ = HJD 2455017.08 was used as the epoch of periastron. The computed orbital parameters are given in Table\ \ref{orbitpar} where they are compared to the results of C05 and A09. The RV curve implied by the solution is shown in Fig.~\ref{rvfit}, bottom. Our values and the ones published by C05 are based only on velocity points from the 
He \begin{small}II\end{small} lines, while A09 applied the velocities of every available H, He \begin{small}I\end{small}, 
and He \begin{small}II\end{small} line, a process which combines lines from two different sources on 
  and  near the O star as mentioned in Section \ref{RV}. 

In general, our computed orbital parameters are close to earlier solutions, but there are some differences. The value of
the computed systemic radial velocity, $V_{\gamma}$, was significantly higher (by 15 to 20 km s$^{-1}$) for each line type (H \begin{small}I\end{small}, He \begin{small}I\end{small} and He \begin{small}II\end{small}) in C05 than what we found. The possibility of a real change in the system RV over a few years is very small, so the cause of the difference is likely due to differences in data analyses.

\subsection{Mass of the compact object}

\begin{figure}
\begin{center}
\leavevmode

\includegraphics[width=8.7cm]{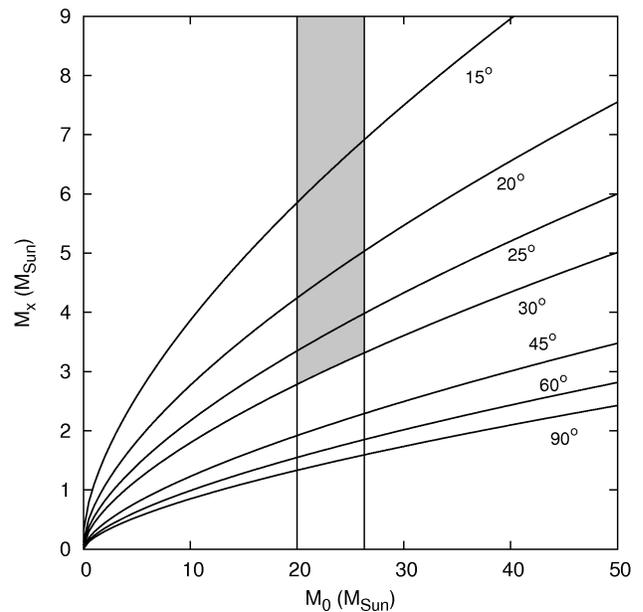}

\end{center}
\caption{Mass constraints of the two components in LS~5039 for different inclinations from the measured mass function. The vertical lines are the limits of 
the O-type stellar companion mass adopted from C05. The gray region represents the possible values of the compact object given that the orbital inclination is less than 30$^{\circ}$.}
\label{massf}
\end{figure}

\begin{figure}
\begin{center}
\leavevmode

\includegraphics[width=8.5cm]{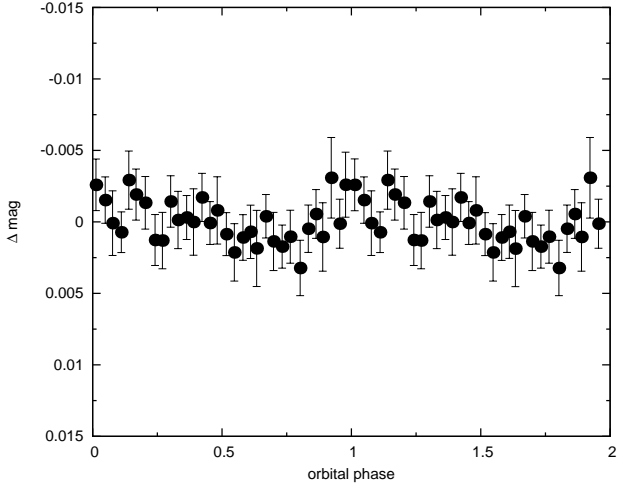}

\end{center}
\caption{Phase-binned folded light curve from the {\em MOST} observations. The error
bars represent 99\% confidence intervals of the mean values {\bf computed from $3 \sigma/\sqrt{n}$ where $\sigma$ is the standard deviation and $n$ is the number of data points in a given phase bin. The {\em MOST} data originally represent fractional change from the mean in terms of flux so a change of sign gives the $\Delta$ mag values shown here that represent the magnitude difference from the mean. This conversion of the {\em MOST} data to $\Delta$ mag allows us to directly compare the data to the simulations shown in Fig.~\ref{modellc}.
}}
\label{mostlc}
\end{figure}

\begin{figure}
\begin{center}
\leavevmode

\hspace*{-1em}
\includegraphics[width=8.7cm]{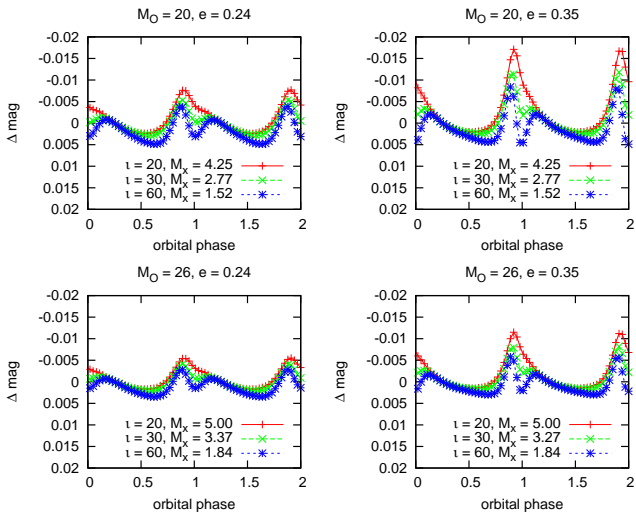}

\end{center}
\caption{{\bf Model light curves. All models have the mass function fixed to 0.0049 M$_{\odot}$. $M_{O}$ is the mass of the primary, $M_{x}$ is the mass of the compact companion. The amplitude of the light curves decrease with increasing system mass and decreasing eccentricity.}}
\label{modellc}
\end{figure}

One of the main goals of our investigation was to obtain stronger constraints on the {\bf mass of the non-stellar companion}. C05
executed detailed light curve simulations using their orbital solution for LS~5039; they found that if the inclination angle 
is around 30$^{\circ}$, then photometric variability caused by the distortion of the primary should be of the order of ~0.01 mag near periastron. If the change in brightness is 0.01 mag or less then the inclination is 30$^{\circ}$ or less. An inclination less than 30$^{\circ}$, in turn, implies that the mass
of the compact object is too high ($>$ 3.0 M$_{\odot}$, C05) to be a neutron star. Lacking the necessary 2-3 mmag 
photometric
accuracy required, they could not check their scenario.

As described in Section \ref{orbit}, our new analysis of the orbital parameters of LS~5039 is based on an independent 
homogeneous radial velocity data set. Our results are in good agreement with the one presented by C05 -- {\bf in particular,  
the values of mass functions, the implications of which are shown in Fig.~\ref{massf}, agree within the uncertainties.}

Photometric data from the {\em MOST} satellite (6649 individual brightness measurements
through 16 days between July 7 and July 23) did not show any variability greater
than a few mmag. To quantify the frequency content of the light curve,
we performed a period analysis of the full dataset using the {\small P{\tiny ERIOD}04} software \citep{Lenz2004}. 
The resulting frequency spectrum does not contain any significant
peak with an amplitude greater than 0.002 mag. Moreover, the orbital period does
not jump out of the noise either. As a different approach, we phased the light
curve with the spectroscopic period and epoch and used 0.03 phase bins to
reveal if there is any optical variability due to orbital motion. {\bf Fig.\
\ref{mostlc} shows the phase-binned folded light curve, which indicates a possible
variability at the level of 2 mmag, with an apparent broad 
minimum at phase $\varphi\sim$ 0.7 -- 0.8.} Although the SNR of the binned light curve is low for detailed speculations, 
we compared the shape of the curve with the EW variations of hydrogen and helium lines {\bf (see Section \ref{starwind}) and with the results of our own light curve simulations.}

{\bf To check and understand the conclusions of the C05 light curve simulations, we used the WD code to do our own simulations. Two sets of simulated light curves were computed: for the first set the mass of the compact object was fixed to 3.0 M$_{\odot}$; for the second set the mass of the compact object was determined from the mass function derived from our radial velocity data as described in Section \ref{orbit}. In each set, the mass of the primary was set to either 20 or 26 M$_{\odot}$, the eccentricity was set to either 0.35 or 0.24, the $V$ band was used to approximate the {\em MOST} band and, inclinations of 60$^{\circ}$, 30$^{\circ}$ and 20$^{\circ}$ were modelled. The results from the second set of simulations, with the mass function fixed to our measured value, are shown in Fig.~\ref{modellc}. For all simulations the primary radius was set to $R_{O} = 9.5$ R$_{\odot}$ and a linear cosine law \citep{Wilson2003} was used for limb darkening effects. The compact object was handled as a point source.

The first set of simulations, with the mass of the compact object fixed at 3.0 M$_{\odot}$, gave light curves that followed the pattern described by C05: the amplitude of the light curve decreased as the inclination decreased. The results with the fixed mass function simulations were different, as seen in Fig.~\ref{modellc}: the amplitude of the light curve decreased with increasing total system mass and decreasing eccentricity but did not decrease with decreasing inclination. For fixed system mass and eccentricity the inclination diagnostic is better given by the light curve shape, especially the dip near phase 0, and not by its amplitude.  A formal curve fit of the {\em MOST} data to the light curves of the $M_{O} = 26$ M$_{\odot}$, $e = 0.24$ case gave the best fit with an inclination of 60$^{\circ}$. However this result must be viewed as preliminary, at best, because the amplitude difference between the different inclination light curves is very small, the binned {\em MOST} data have considerable scatter at the millimag level, and the effect of the stellar wind and accretion on the light curve are not modelled in the WD code.}

We note that some studies suggest the mass of an O6.5V type star to be around 28 -- 29 $M_{\odot}$ 
\citep{Martins2005} rather than {\bf 20 -- 26 $M_{\odot}$ used in C05. }
{\bf In the case of a heavier primary, the mass of the compact object would also be larger and the $M_{x}$ masses shown in Fig.~\ref{modellc} would be greater, strengthening the argument for the black hole nature of the secondary compact star, even at relatively higher inclinations.}

\begin{figure}
\begin{center}
\leavevmode

\includegraphics[width=8.5cm]{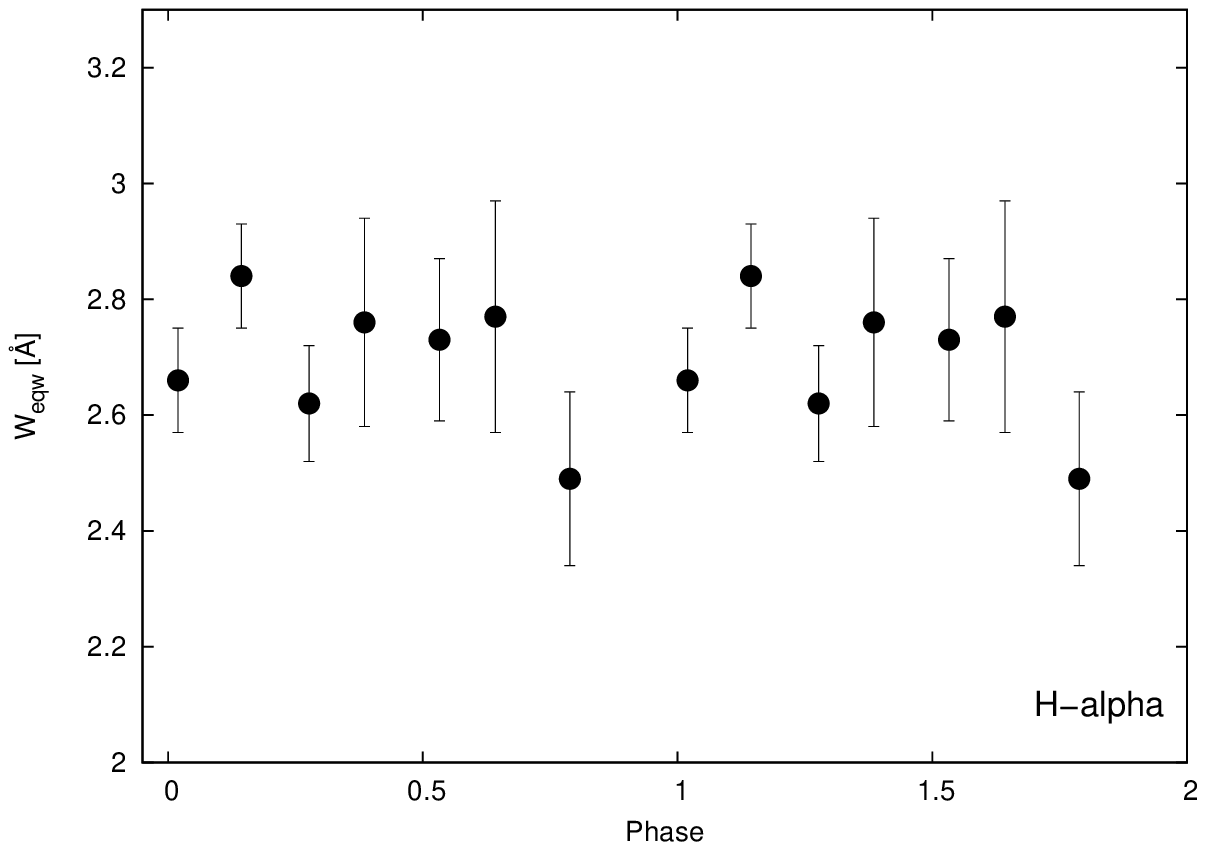}
\includegraphics[width=8.5cm]{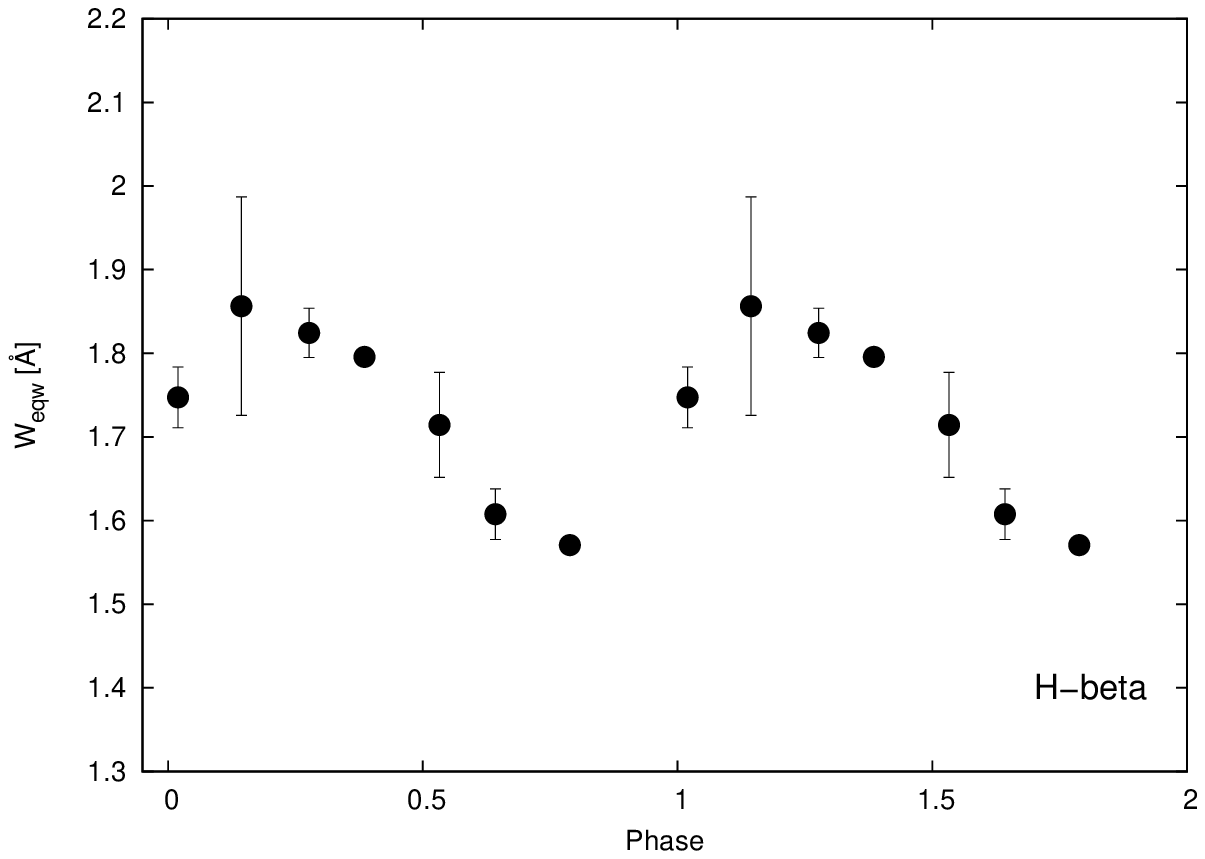}
\includegraphics[width=8.5cm]{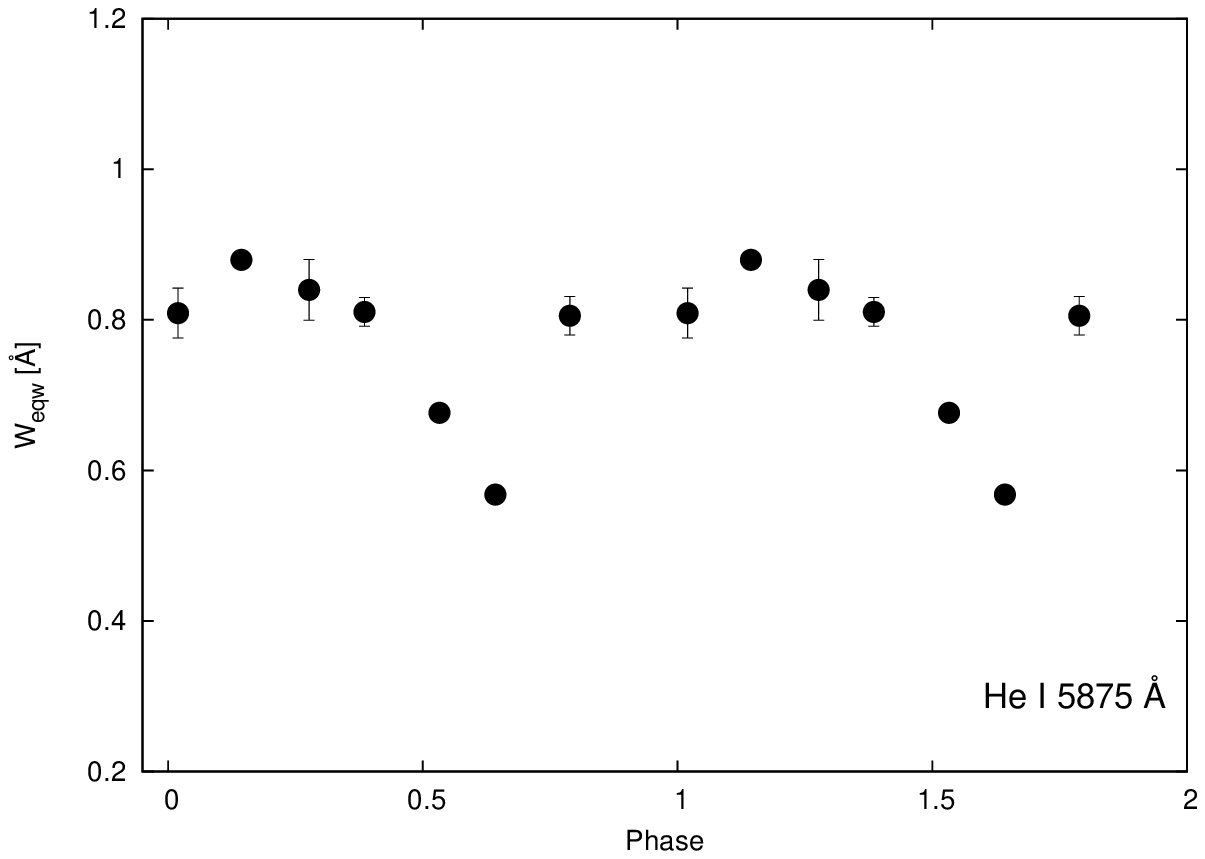}

\end{center}
\caption{Variability of equivalent widths of H$\alpha$, H$\beta$ and He I $\lambda$5875 absorption lines during the orbit.
The mean values and the given errorbars of points were calculated from results {\bf we got} from three different methods (direct
measurement with {\small IRAF} {\tt splot} task, and fitting Gaussian or Lorentzian curve to the line profile) {\bf used} to
determine EWs of lines.}
\label{abs_EW}
\end{figure}

\subsection{Stellar wind}\label{starwind}

To infer the mass loss rate of the O-type star and the properties of the circumstellar matter, we determined the 
equivalent widths of each H and He line and also their variability during the orbit, which could be good indicators of 
physical processes taking place in the stellar wind. As mentioned in Section 3.2, we also measured the EWs of some
other lines to check the value of interstellar reddening given in the literature. We made the measurements on the 
average of
spectra summed to one hour long exposure times, but we plotted the daily average values of the EWs on final 
diagrams, shown in Fig.~\ref{abs_EW}, to better see the trends. 

For the H$\alpha$ line we found that EW changes from 2.50 to 2.85 \AA\ over an orbital phase (see 
Fig.~\ref{abs_EW}). The average value of 2.70 \AA\ agrees very well, within the uncertainties, with the 
result of C05 (2.8 $\pm$ 0.1 \AA), except they found that the value was stable during the orbit of the binary (which 
could possibly be explained by the relatively low resolution of their spectra). Our result is also consistent with the EW values 
measured by others over the last ten years \citep{Bosch-Ramon2007}. Using the method of \citet{Puls1996} we 
estimated the mass loss rate from the EW of the H$\alpha$ line. To do these calculations we adopted the following 
parameters: $R_{\rm O}$ = 9.3$^{+0.7}_{-0.6}$, R$_{\odot}$ and $T_{\rm eff}$ = 39,000 $\pm$ 1000 K for the O-type star (C05); 
a terminal wind
velocity of $V_{\infty}$ = 2440 $\pm$ 190 km s$^{-1}$ with a wind velocity law exponent of $\beta$ = 0.8 (M04).
We found that the mass-loss rate of the stellar companion is around 3.7 $\times$ 10$^{-7}$ M$_{\odot}$ yr$^{-1}$ from 
the strongest absorption, 
  which corresponds to the lower limit, and 4.8 $\times$ 10$^{-7}$ M$_{\odot}$ yr$^{-1}$ for the upper limit. 
 These values 
 are consistent with the mass loss rates 
 of $\approx 3.7 \times 10^{-7}$ M$_{\odot}$ yr$^{-1}$ for the low state (strong absorption) 
 and $\approx 7.5 \times 10^{-7}$ M$_{\odot}$ yr$^{-1}$ for the high state obtained by C05 (see also M04).  

We found two lines (H$\beta$ and He \begin{small}I\end{small} $\lambda$5875) showing significant changes during the orbit. 
The lowest absorption for the H$\beta$ line occurs around $\varphi\sim$ 0.75, and at 
$\varphi\sim$ 0.65  for the He \begin{small}I\end{small} line (Fig.\ \ref{abs_EW}), close to the expected phase 
$\varphi\sim$ 0.7 when the compact object is between us and the stellar companion (inferior conjunction). {\bf The lower EW at inferior conjunction implies an increased emission strength likely due to the focusing of the stellar wind toward the compact object. Models of the wind flow in the system will need to take that focusing into account, especially those that model a pulsar wind/stellar wind interaction.}
We carried out a simple correlation analysis between these EW changes and the {\em MOST} light curve {\bf (rebinned to the EW bin size)}, and found
the correlation coefficients, $r$, to be 0.52, 0.70 and 0.52 for the H$\alpha$, H$\beta$ and He \begin{small}I\end{small} 
$\lambda$5875 lines, respectively. This result suggests the possibility of real modulations over the orbital period, but further
studies are necessary.

Note that our data only show smooth orbital modulation 
  in the H$\alpha$, H${\beta}$ and He {\small I} absorption lines and in the emission components of the H Balmer lines. 
The lines do not show evidence of dense clump-like condensates in the stellar wind.

\subsection{Some remarks}\label{remarks}

The two main findings from our work are that 
  the orbit has a significantly lower eccentricity than previously thought
  and that the {\bf total mass of the system is toward the higher end of previous estimates, when the variability of the {\em MOST} data is considered. We were not able to resolve the nature of the compact object, between neutron star and black hole, in spite of simulations by C05 that postulated that a low amplitude light curve would be associated with a low inclination and a black hole.}  
The eccentricity $e\approx 0.4$ obtained by M04 put LS~5039 as the system with the most eccentric orbit 
   among X-ray binaries with an O  donor star. 
Although the value was later shown to be lower ($e\approx 0.35$, C05 and A09), 
   the eccentricity would still be high enough to have some consequences 
   on the dynamical and radiative properties, and on the evolutionary history, of the system.  
{\bf The observed TeV spectrum is consistent with inverse Compton scattering from electrons distant from the compact object either in a pulsar wind/stellar wind shock if the compact object is a neutron star, or in a jet if the compact object is a black hole. The lack of accretion features in the X-ray spectrum favors the pulsar model unless the bulk of the accretion power is released as kinetic energy in jet outflow instead of thermally from an accretion disk \citep{Takahashi2009}. In either case the keV X-rays are hypothesised to originate as synchrotron emission from the electrons responsible for the inverse Compton TeV emission. This is in contrast to the canonical view of X-ray emission from X-ray binaries.}
In the canonical view, the keV X-rays from X-ray binaries with an O star  
 are powered by the accretion of gas either in a wind focused by the Roche-lobe of the donor star 
  or captured directly from the wind out-flowing from the donor star. 
In the focused wind scenario, an accretion disk is expected to form around the compact star, 
  similar to that in systems with Roche-lobe filling mass-transfer,  
  as the accreting gas carries substantial specific angular momentum. 
In the direct wind capture scenario, provided that the specific angular momentum of the gas is sufficiently small, 
  an extensive accretion disk might not be formed 
   and the accretion inflow would be practically radial and resemble that of a Bondi-Hoyle flow. 
A {\bf lower orbital} eccentricity would give a lower specific angular momentum in the wind material 
   that is swept up by the accreting star. 
So far observations [e.g. X-ray observations by {\it Chandra} and {\it XMM-Newton}  
   \citep{Martocchia2005,Bosch-Ramon2007}] have not shown evidence of an accretion disk in LS~5039.    
If LS~5039 has an extremely high orbital eccentricity as originally measured,   
  the lack of an accretion disk around its compact star would need certain non-trivial explanations  
  (cf. the situation in the Be X-ray binaries). 
The lower eccentricity may ease the situation somewhat,  
  allowing radial gas inflow for a substantial distance before reaching the compact accretor.   
Then, if there is an accretion disk, it would not be expected to be a large and dense disk 
because such a disk would generate significant thermal soft keV X-rays, 
   which are not detected in the {\it Chandra}, {\it XMM-Newton} and other X-ray observations.  
Our smaller value of eccentricity reduces the discrepancy 
  between the low observed X-ray variability and 
  what would be expected from  wind accretion with Bondi-Hoyle like radial inflow
  [\citet{Bondi1944}, \citet{Bosch-Ramon2005}, C05].

Known black-hole high-mass X-ray binaries tend to have small orbital eccentricity, 
  unlike the X-ray pulsars (see \citep{Liu2005,Liu2006,Liu2007}).  
X-ray binaries with a black hole and an O donor star are very rare, 
  and Cyg X-1 is the currently only known system in the Milky Way.  
The confirmation of LS~5039 as a black-hole high-mass X-ray binary 
   in which a massive O donor star and a black hole revolve around each other in an eccentric orbit 
   {\bf would have} important implications on how such systems are formed and how massive binaries evolve.    
In LS~5039, the compact star's progenitor, at a certain stage, should be more massive 
   than the current O donor star, 
   otherwise it would not have evolved to form the compact star.   
O stars have very short life spans ($\sim$ a few Myr), 
  thus LS~5039 as {\bf an X-ray binary} must be younger than a few million years.  
This is supported by the fact that LS~5039 has a highly eccentric binary orbit, 
  which has yet to be {\bf circularized} from a presumed recent supernova event.  
However, it is unclear  
  whether the progenitor of the O star or the progenitor of the compact star  
  had the larger initial mass, 
  {\bf as a rapid} evolution of the progenitor of the compact star could well be triggered 
  by a mass transfer process,  
  in which the system was compact enough to allow the progenitor of the current O star to overfill its Roche-lobe and transfer material to the progenitor of the compact star. 
(For more on the evolution of very massive binaries, see e.g. \citet{Dalton1995,VanBever2003,Dionne2006}.)
If LS~5039 was indeed formed through this mass-transfer channel, 
  {\bf a finding that it contains a black hole} among the fewer than 200 known high-mass X-ray binaries in the Milky Way  
  would imply that a black-hole high-mass X-ray might be formed in close massive binaries more easily than previously thought. 
Young stellar clusters in star-forming galaxies may well be populated by LS~5039-type sources 
   and are potentially Gev-TeV $\gamma$-ray sources.

\section{Conclusions}\label{four}

 Our simultaneous optical photometry from the {\em MOST} space telescope and high resolution echelle 
   optical spectroscopy from the ANU 2.3m Telescope 
   have put {\bf further} constraints on the orbital parameters of the LS~5039 system. 
In particular we obtained a mass function $f(m) \approx0.0049 \pm 0.0006$ M$_{\odot}$
  and an orbital eccentricity $e=0.24\pm0.08$. 
The maximum photometric variation of LS 5039 in the {\em MOST} light curve was 2 mmag. 
{\bf The low photometric variation is consistent with the lower orbital eccentricity of 0.24, as opposed to the value of 0.35 found by others, and it is consistent with a higher overall system mass. However, we cannot determine the inclination on the basis of the low photometric amplitude. Our light curve simulations imply that the mass of the compact object is at least 1.8 M$_{\odot}$ based on an inclination of 60$^{\circ}$ or less.}   
Our value for the eccentricity of 0.24$\pm$0.08 is a little smaller than previous determinations. 
The lower eccentricity implies that the wind material that is captured and falls into the Roche lobe of the compact star 
   has a lower specific angular momentum. 
Thus it may not lead to the formation of a large-scale optically thick accretion disk, 
   and the accretion inflow is practically radial resembling that of a Bondi-Hoyle flow. {\bf This radial inflow would explain the lack of an accretion disk signature in keV data and should be considered in models of mass transfer for either a non-accreting pulsar or a black hole.} 
Finally from EW measurements of the H$\alpha$ line, 
   we derived that the mass loss rate from the O-type primary through stellar wind 
   is 3.7 to 4.8 $\times 10^{-7}$ M$_{\odot}$ yr$^{-1}$, 
   similar to values obtained by other workers.  
Our observations do not show evidence of dense clumps in the stellar wind.

\section*{Acknowledgments} 

This work has been supported by the Australian Research Council, the University
of Sydney, the Hungarian OTKA Grants K76816 and MB0C 81013, and the ``Lend\"ulet'' Young
Researchers' Program of the Hungarian Academy of Sciences. 
KW's visits to Sydney University were supported by the University of Sydney International Visiting Fellowship.

\label{lastpage}

\end{document}